\documentclass[10pt,conference]{IEEEtran}
\IEEEoverridecommandlockouts

\usepackage[T1]{fontenc}
\usepackage[latin9]{inputenc}
\usepackage{graphicx,kantlipsum,setspace}
\usepackage{calc}

\usepackage{amsthm}
\usepackage{pdfpages}
\usepackage{lipsum}
\usepackage{amsmath}
\DeclareMathOperator{\diag}{diag}
\usepackage{mathtools}
\usepackage{venndiagram}
\usepackage{tikz-cd}
\usepackage{amssymb}
\usepackage{tikz}
\usepackage{subcaption}
\usepackage{setspace}
\usepackage{caption}
\usepackage{setspace}
\usepackage[left=0.625in,top=0.73in,bottom=0.98in,right=0.625in]{geometry}

\usepackage[]{}

\newlength{\myheight}
\makeatletter

\DeclareRobustCommand*{\lyxarrow}{%
\@ifstar
{\leavevmode\,$\triangleleft$\,\allowbreak}
{\leavevmode\,$\triangleright$\,\allowbreak}}

\theoremstyle{plain}
\newtheorem{thm}{\protect\theoremname}
\theoremstyle{plain}

\makeatother

\providecommand{\lemmaname}{Lemma}
\providecommand{\theoremname}{Theorem}

\begin{document}
\title{Task-Oriented Communication with Out-of-Distribution Detection: An Information Bottleneck Framework\vspace*{-8mm}}

\author{\IEEEauthorblockN{Hongru~Li, Wentao~Yu, Hengtao He, Jiawei~Shao, Shenghui~Song, \\Jun Zhang,~\textit{Fellow,~IEEE,} and Khaled B.~Letaief,~\textit{Fellow,~IEEE}\\\IEEEauthorblockA{Dept. of Electronic and Computer Engineering,
The Hong Kong University of Science and Technology, Hong Kong\\
Email: \{hlidm, wyuaq\}@connect.ust.hk, eehthe@ust.hk, jiawei.shao@connect.ust.hk, \{eeshsong, eejzhang, eekhaled\}@ust.hk}}\\ \vspace*{-12mm}}

\IEEEspecialpapernotice{}
\maketitle
\begin{abstract}
\textit{Task-oriented communication} is an emerging paradigm for next-generation communication networks, which extracts and transmits task-relevant information, instead of raw data, for downstream applications. Most existing deep learning (DL)-based task-oriented communication systems adopt a \textit{closed-world} scenario, assuming either the same data distribution for training and testing, or the system could have access to a large out-of-distribution (OoD) dataset for retraining. However, in practical \textit{open-world} scenarios, task-oriented communication systems need to handle unknown OoD data. Under such circumstances, the powerful approximation ability of learning methods may force the task-oriented communication systems to overfit the training data (i.e., in-distribution data) and provide overconfident judgments when encountering OoD data. Based on the information bottleneck (IB) framework, we propose a class conditional IB (CCIB) approach to address this problem in this paper, supported by information-theoretical insights. The idea is to extract distinguishable features from in-distribution data while keeping their compactness and informativeness. This is achieved by imposing the class conditional latent prior distribution and enforcing the latent of different classes to be far away from each other. Simulation results shall demonstrate that the proposed approach detects OoD data more efficiently than the baselines and state-of-the-art approaches, without compromising the rate-distortion tradeoff.
\end{abstract}


\IEEEpeerreviewmaketitle{}

\section{Introduction}
Recent years have witnessed significant achievements of artificial intelligence (AI) in numerous fields, such as smart healthcare, industrial AI and autonomous vehicles. One of the latest advancements is Edge AI, which deploys AI applications directly onto edge devices such as smartphones, embedded systems and other Internet of things devices. Edge AI has brought about a paradigm shift in how data is processed and transmitted for AI applications~\cite{letaief2019roadmap}, as it provides low-latency and reliable local inference. However, the heavy traffic for Edge AI will induce tremendous communication overhead~\cite{letaief2021edge}. To address this issue, next-generation communication systems are expected to shift from being \textit{data-oriented} to \textit{task-oriented}. This shift aims to reduce the communication overhead and inference latency by extracting and delivering only the task-relevant information for downstream tasks.

\begin{figure*}[ht]
\centering
\includegraphics[width=0.75\linewidth]{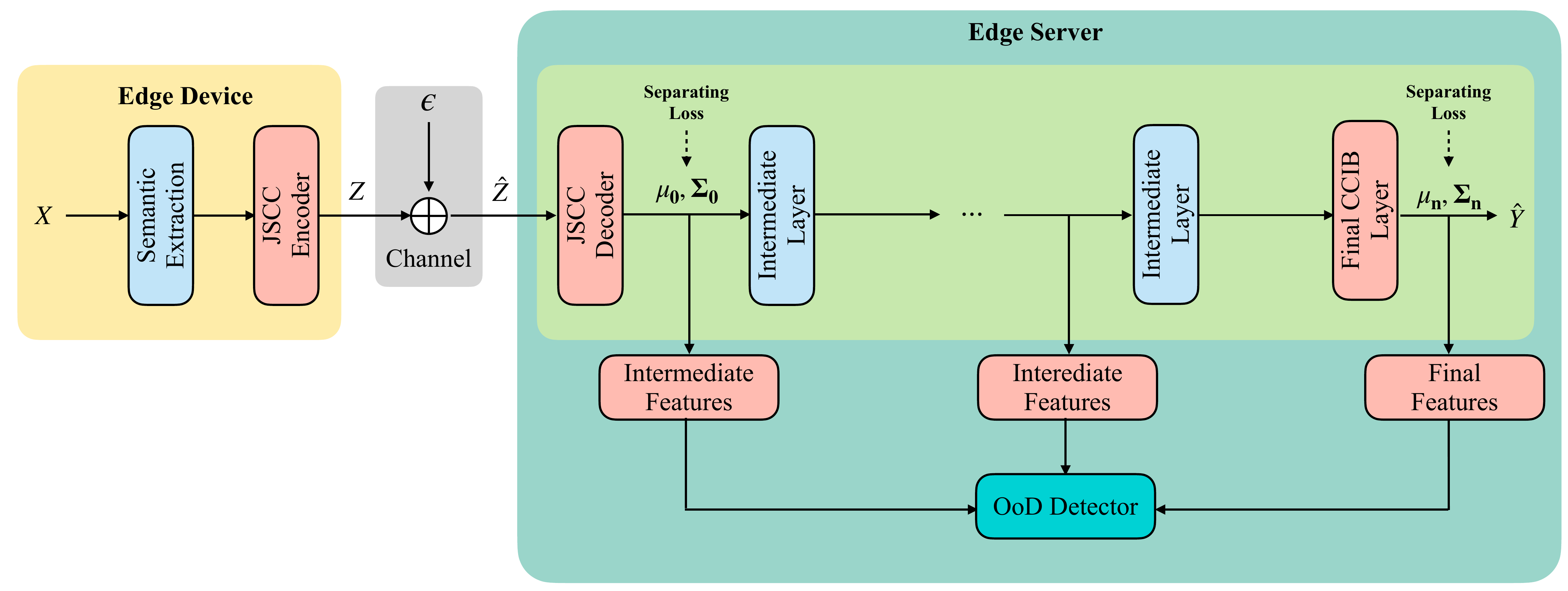}
\caption{The framework of the proposed CCIB-based task-oriented communication system.}
\label{fig:pic1}
\end{figure*}
To perform task-relevant information extraction and transmission, two main schemes are typically used: separate source-channel coding (SSCC) and joint source-channel coding (JSCC). Shannon's theorem proved that the SSCC scheme is optimal only in the infinite blocklength regimes~\cite{shannon1959coding}. In the finite blocklength scenarios, the SSCC scheme has significant performance loss, and the JSCC scheme can achieve more efficient and reliable coding performance. This is because JSCC can take advantage of the inherent relationships between the source and channel coding. Recently, there has been a growing interest in deep learning (DL)-based JSCC design, spanning a range of applications such as image-based~\cite{bourtsoulatze2019deep}, text-based~\cite{xie2021deep}, and video-based~\cite{jiang2022wireless} task-oriented communication systems. However, the joint optimization between the communication overhead and the task execution performance was not considered by above methods. To address this, the authors of~\cite{shao2021learning} utilized the information bottleneck (IB) framework to explicitly formulate the above-mentioned joint optimization as a rate-distortion optimization problem. The IB framework was later extended to multi-device cooperative edge inference in~\cite{shao2022multi} and sequence data processing in~\cite{shao2022task}. The objective of the IB framework is to maximize the mutual information between the latent representation and label (to minimize \textit{distortion}), while minimizing the mutual information between the latent representation and input data (to minimize \textit{rate}).

Although the aforementioned DL-based task-oriented communication systems have achieved great successes, the issue of distribution shift, i.e., the distribution of the testing data is significantly different from that of the training data, has not been well tackled for practical deployment. Specifically, in the \textit{closed-world} assumption, the data distribution for training and testing are considered the same or the systems could access a large OoD dataset for adaptation. However, with the more realistic \textit{open-world} assumption, these learning-based methods may provide overconfident judgments when exposed to OoD data that was not seen during the training phase. This can significantly compromise the inference performance or even lead to unacceptable consequences, especially for task-oriented communications. There have been some works tackling this issue. In~\cite{dai2022adaptive}, the authors proposed a test-time update scheme, which updates the model parameters for each test instance. However, this method may lead to an extremely high communication overhead due to the back-propagation between the transceivers. In~\cite{zhang2022deep}, the authors proposed a cycle generative adversarial network-based model to convert the observed OoD data to similar in-distribution (ID) data. However, this approach assumed that the system can foresee the OoD dataset for adaptation, but did not provide any OoD detection mechanism to determine the data that should be converted. In fact, predicting the distribution of incoming OoD data is difficult and it is more troublesome to have a large OoD dataset for retraining and ensuring the compactness and informativeness of the extracted features for edge inference. Thus, there is an urgent need for a method to \textit{detect} the OoD data, and advocate \textit{detection} for \textit{collection} and \textit{adaptation}.

To solve these issues, we propose a class conditional information bottleneck (CCIB)-based approach for extracting distinguishable, compact, and informative features for efficient OoD detection. To the best of the authors' knowledge, this is the first work to address the OoD detection problem in task-oriented communication systems. Specifically, we extend the IB framework to a CCIB approach. The proposed approach imposes class conditional latent priors and applies a contrastive learning method to ensure the separation of the latent space corresponding to each data class. The ID features extracted by this approach are more distinguishable than the OoD features that are not associated with any ID label. Additionally, we leverage the explicit rate-distortion optimization objective in IB to keep the extracted features compact and informative. The simulation results over image classification tasks show that the proposed method can extract more distinguishable features than the baselines while keeping the features compact and informative for detecting the OoD data.

\section{System Model and Problem Formulation}
As shown in Fig. \ref{fig:pic1}, we consider a task-oriented communication system for edge-device co-inference with OoD detection, where the main task is image classification. The system consists of an edge device equipped with a semantic feature extraction module and a JSCC encoder, as well as an edge server with a JSCC decoder. During the training phase, $N$ training data points $\{\mathbf{x}, y\}^N_{n=1}$ from source domain are utilized as ID data to optimize system parameters. During the testing phase, the system encounters both ID and OoD data. The on-device network extracts task-related features, denoted as $Z$, and transmits them to the server. The server network then utilizes the noise-corrupted features, represented by $\hat{Z}$, to produce the main task results, $\hat{Y}$, while concurrently identifying OoD data. 
For simplicity, we consider an additive white Gaussian noise (AWGN) channel with noise $\mathbf{\epsilon}\sim\mathcal{N}(0,\sigma^2 \mathbf{I})$. The channel condition is modeled in terms of the peak signal-to-noise ratio (PSNR), given by
\begin{equation}
    \text{PSNR}=10\log\frac{P}{\sigma^2},
\end{equation}
where $P$ is the maximum transmit power of each feature dimension and $\sigma^2$ refers to the noise power.

In the previous works, the IB-based systems have been shown capable of achieving a better rate-distortion tradeoff than other schemes~\cite{shao2021learning}. The underlying probabilistic graphical model of the IB method is shown below
\begin{equation}\label{eq:original_graph}
\begin{tikzcd}[row sep=tiny]
Y&X\arrow[l]\arrow[r]&Z\arrow[r]&\hat{Z}\arrow[r]&\hat{Y},
\end{tikzcd}
\end{equation}
and the rate-distortion tradeoff is optimized by 
\begin{equation}\label{eq:original_opt}
\mathcal{L}_{IB}\triangleq \underbrace{\beta I(X;\hat{Z})}_{rate}- \underbrace{I(Y;\hat{Z})}_{distortion}.
\end{equation}

In (\ref{eq:original_opt}), the mutual information between the input $X$ and the received latent $\hat{Z}$, denoted as $I(X;\hat{Z})$, represents the preserved information of $\hat{Z}$ given $X$. On the other hand, the mutual information between the label $Y$ and $\hat{Z}$, denoted as $I(Y;\hat{Z})$, measures how informative the received latent is for reducing the uncertainty in predicting the label. 

However, the optimization of the IB objective involves an intractable distribution, i.e., the prior latent distribution $p(\hat{z})$, which requires the use of a tractable surrogate prior distribution. In the previous works, the surrogate prior distribution of the latent variable was assumed to be independent of the joint distribution of the input and its corresponding label. Specifically, previous works assumed either a fixed log-uniform prior~\cite{shao2021learning} or a centered isotropic Gaussian prior~\cite{shao2022multi}. These assumptions make the system insensitive to detect the OoD data because both the ID and OoD data are mapped to the same latent distribution without being associated with an ID label.



To make this framework more efficient in detecting the OoD data, we need to address two issues. The first one is how to make the on-device network extract distinguishable features while maintaining compactness and informativeness. The second is how to make the server network sensitive to the noise-corrupted features of the OoD data and effective in detecting the OoD data based on them.

\section{A CCIB Approach For Task-Oriented Communication}\label{system model}


In this section, we introduce a novel CCIB approach for task-oriented communication systems, which enhances the distinguishability of ID features against the OoD features, while maintaining the compactness and informativeness of ID features. This is achieved by assigning a class-conditional latent prior and enforcing separation between the latents of different classes. We also exploit the variational approximation to derive a tractable surrogate objective for the intractable distribution involved in the optimization of the CCIB framework.

\begin{figure}[t]
    \centering
    \begin{subfigure}{0.7\linewidth}
        \centering
        \includegraphics[width=0.9\linewidth]{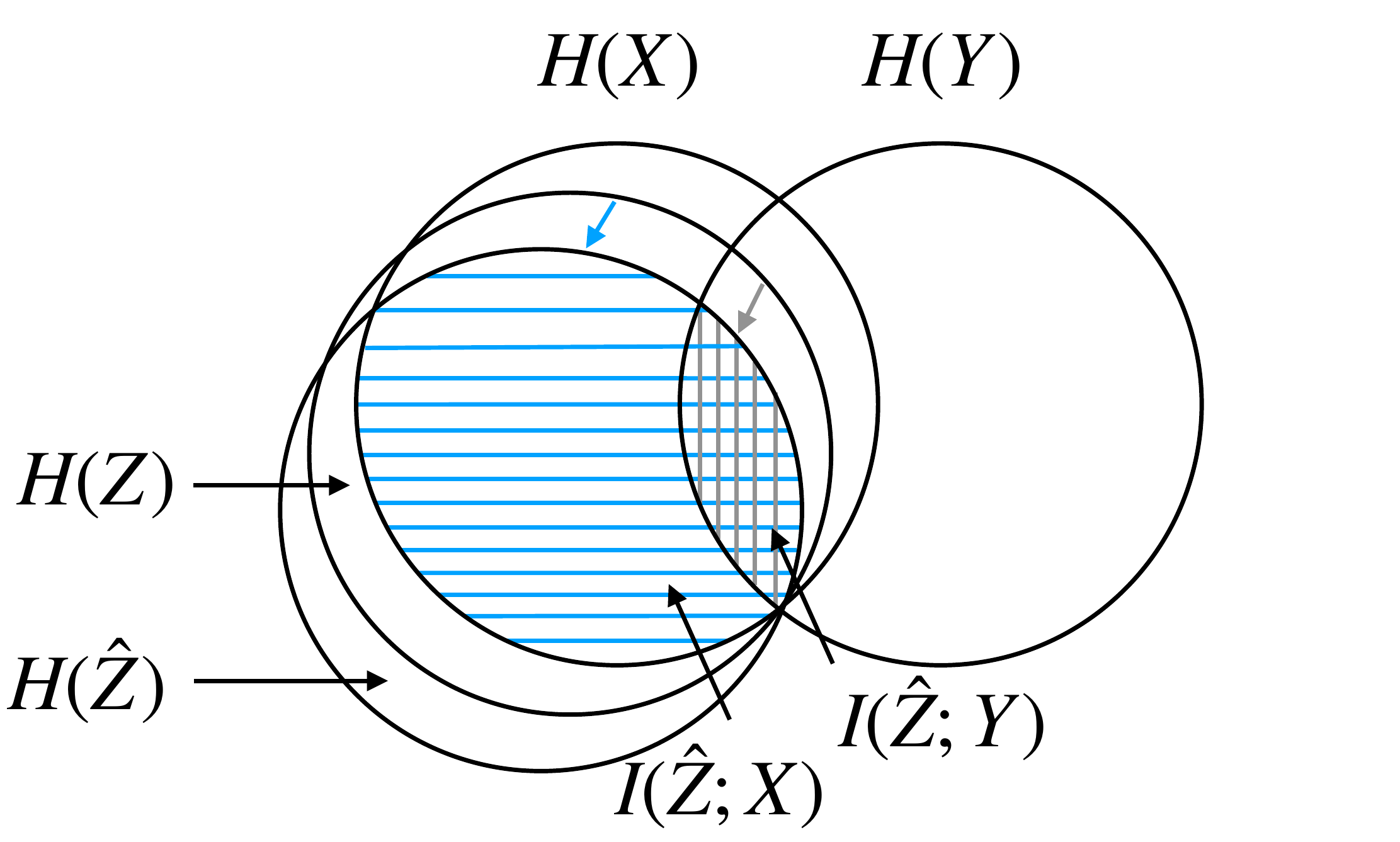}
        \caption{Objective of IB}
        \label{fig:ib_venn}
    \end{subfigure}
    \quad\quad\quad\quad\quad\quad\quad\quad\quad
    
    \begin{subfigure}{0.7\linewidth}
        \centering
        \includegraphics[width=0.9\linewidth]{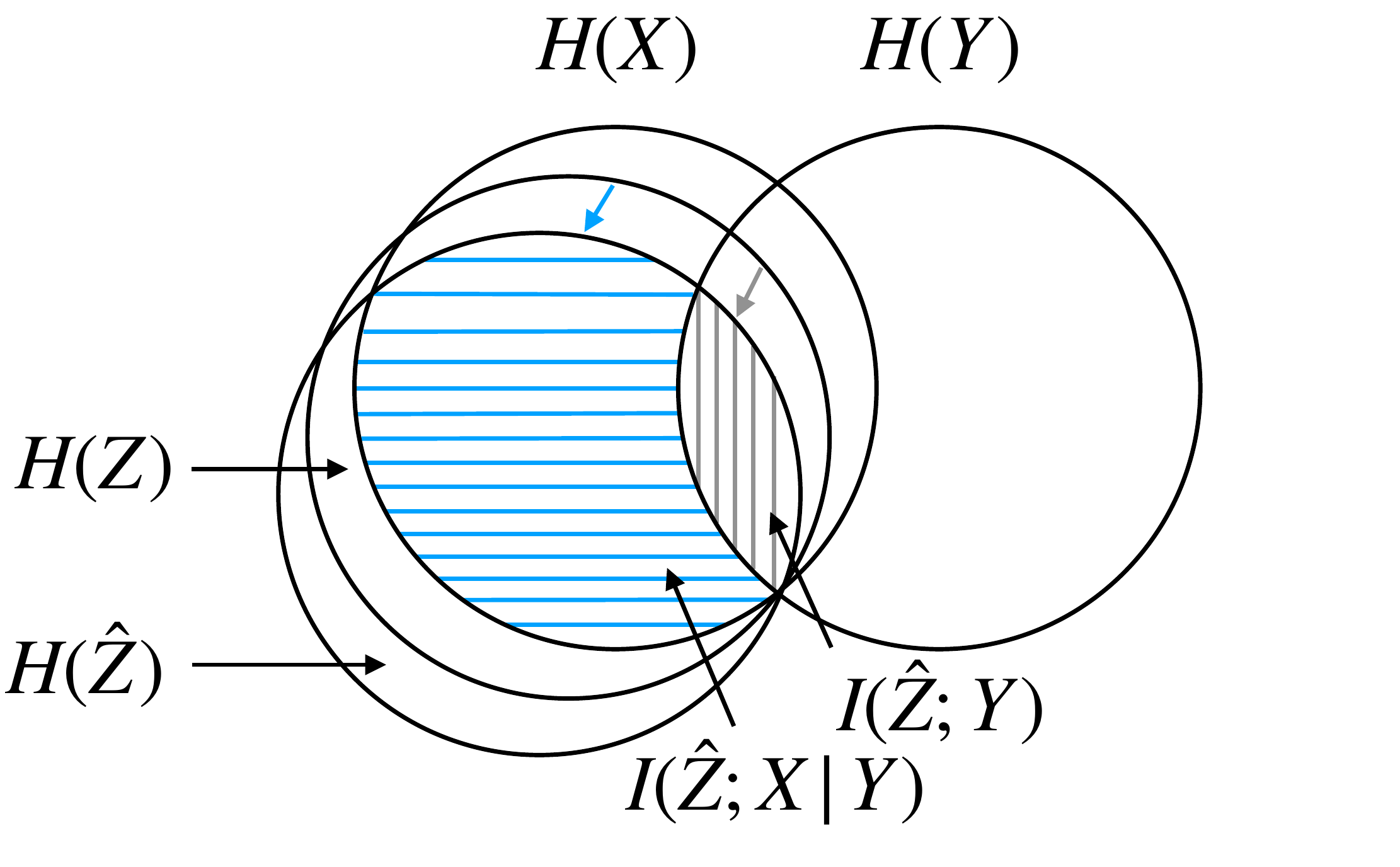}
        \caption{Objective of CCIB}
        \label{fig:ccib_venn}
    \end{subfigure}
    \caption{The Venn diagrams of IB and CCIB show optimization (maximum and minimum) areas in their objectives. Blue lines: the area being minimized by both objectives. Gray lines: the area being maximized by both objective. The colored arrows indicate the information loss because of the channel noise. \textbf{IB}: The term to be maximized, i.e., $I(\hat{Z};Y)$, is a subset of the term to be minimized, namely, $I(\hat{Z};X)$. \textbf{CCIB}: The term to be maximized, i.e., $I(\hat{Z};Y)$, is no longer a subset of the term to be minimized, namely, $I(\hat{Z};X|Y)$.\\ \vspace*{-8mm}}

    \label{fig:venn}
\end{figure}
\addtolength{\footskip}{0.04in}
\subsection{Proposed CCIB Scheme}
To overcome the fixed prior limitation for OoD detection in the IB framework, we propose a CCIB method which incorporates the class conditional prior. This is achieved by revising the optimization objective in IB (\ref{eq:original_opt}) to a class conditional form, which is given by
\begin{equation}\label{eq:proposed_opt}
\mathcal{L}_{CCIB}\triangleq\underbrace{\beta I(X;\hat{Z}|Y)}_{redundancy}- \underbrace{I(Y;\hat{Z})}_{distortion}.
\end{equation}
The first term in (\ref{eq:proposed_opt}) aims at minimizing the mutual information between the input $X$ and the received latent $\hat{Z}$ conditional on the label $Y$. 
Upon receiving the latent feature $\hat{Z}_o$ from an OoD instance $x_o$, the likelihood of this latent feature tends to exhibit a high entropy\footnote{This means OoD features have almost equal likelihood with respect to all ID labels.} with respect to all ID labels, as there is no associated label $Y_s$ for the OoD instance.
The second term aims to maximize the mutual information of the received latent $\hat{Z}$ and the label $Y$. 
By incorporating the label information to guide the latent distribution, this approach builds the connections between ID inputs and their extracted features with unique prior distributions, which can help capture the specific ID latent patterns for each ID label.

To better illustrate the relation between the objectives of IB and CCIB, we present the Venn diagrams in Fig. \ref{fig:venn}. In both frameworks, $Y$ and $Z$ are conditionally independent given $X$ due to the probabilistic graphical model: $Y\leftarrow X\rightarrow Z\rightarrow \hat{Z}$. This means that the information about $Y$ contained in $Z$ or $\hat{Z}$ is entirely derived from $X$. However, the Venn diagram of IB shows that the term to be minimized, i.e., $I(\hat{Z};X)$, is a subset of the term to be maximized, namely, $I(\hat{Z};Y)$, which may lead to conflicts during optimization~\cite{fischer2020conditional}. By revising the term to be minimized from $I(\hat{Z};X)$ into $I(\hat{Z};X|Y)$, the objective of CCIB allows to optimize the mutual information without conflicts; meanwhile, we can utilize the information of $Y$ to guide the latent distribution of $\hat{Z}$. This means that each class has a unique latent prior distribution, which enables us to extract more distinguishable ID features, as the latent distribution of the OoD data is not associated with any ID label.

In the previous analysis, we highlighted the benefits of the objective of CCIB. However, like IB~\cite{shao2021learning}, CCIB faces the challenges caused by the high-dimensional integrals in the latent prior $p(\mathbf{\hat{z}})$ and the label posterior distribution $p(y|\mathbf{\hat{z}})$, which are computationally prohibitive. To address this issue, we use the variational approximation method to derive a tractable variational upper bound $\mathcal{L}_{VCCIB}$ in~\textbf{Theorem \ref{thm:vccib}}. By optimizing this surrogate upper bound $\mathcal{L}_{VCCIB}$ instead, we can avoid the above-mentioned intractability issue.

\begin{thm}\label{thm:vccib} The variational upper bound of (\ref{eq:proposed_opt}) is given by
\begin{equation}
\begin{aligned}
\mathcal{L}_{CCIB}&\leq \beta\int_{\mathcal{X,Y,\hat{Z}}}p(\mathbf{x},y)D_{KL}(p_{\varphi}(\mathbf{\hat{z}}|\mathbf{x})||r(\mathbf{\hat{z}}|y))d\mathbf{x}dyd\mathbf{\hat{z}}\\
&\underbrace{\quad -\int_{\mathcal{X,Y,\hat{Z}}}p(\mathbf{x},y)p_{\varphi}(\mathbf{\hat{z}}|\mathbf{x})\log q_{\psi}(y|\mathbf{\hat{z}})d\mathbf{x}dyd\mathbf{\hat{z}}}_{\mathcal{L}_{VCCIB}(\varphi,\psi)},
\end{aligned}
\end{equation}
where $p_{\varphi}(\mathbf{\hat{z}}|\mathbf{x})$ is the conditional latent distribution given $(\mathbf{x}, y)$. $r(\mathbf{\hat{z}}|y)$ and $q_{\psi}(y|\mathbf{\hat{z}})$ are the variational approximation of the conditional latent prior $p(\mathbf{\hat{z}}|y)$ and $p(y|\mathbf{\hat{z}})$, repectively. Here, $\varphi\text{ and }\psi$ are the parameters of the distribution family of $p_{\varphi}(\hat{\mathbf{z}}|\mathbf{x})\text{ and }q_{\psi}(y|\mathbf{\hat{z}})$, and $D_{KL}(p(\mathbf{x})||q(\mathbf{x})))=\int_{\mathcal{X}}p(\mathbf{x})\log\frac{p(\mathbf{x})}{q(\mathbf{x})}d\mathbf{x}$ denotes the Kullback-Leibler divergence between $p(\mathbf{x})$ and $q(\mathbf{x})$.
\end{thm}
\begin{IEEEproof} Please refer to Appendix \ref{FirstAppendix}.
\end{IEEEproof}

\subsection{DNN Parameterization and Contrastive Learning-Based Separation Loss}
In the proposed task-oriented communication system, we utilize deep neural networks (DNNs) to parameterize the distribution family by exploiting the universal approximation property of DNNs. Specifically, we model $\varphi$ and $\psi$ by the parameters of the on-device network and the sever-based network, respectively. Similar to previous works~\cite{shao2021learning},~\cite{alemi2016deep}, we model the conditional prior $r(\mathbf{\hat{z}}|y)$ and the posterior latent distribution $p_{\varphi}(\mathbf{\hat{z}}|\mathbf{x}_n)$ as a Gaussian distribution, i.e., $p_{\varphi}(\mathbf{\hat{z}}|\mathbf{x})=\mathcal{N}(\mathbf{\hat{z}}|\boldsymbol{\mu},\boldsymbol{\Sigma})$, and $r(\mathbf{\hat{z}}|y=c)=\mathcal{N}(\mathbf{\hat{z}}|\boldsymbol{\mu}_{c},\boldsymbol{\Sigma}_{c})$, where $\boldsymbol{\mu} = (\mu_1,\dots,\mu_m)\in \mathbb{R}^m$ and $\boldsymbol{\mu}_c = (\mu_{1,c},\dots,\mu_{m,c})\in \mathbb{R}^m$ denote the mean vectors, while $
\boldsymbol{\Sigma}=\diag\{\sigma^2_1,\dots,\sigma^2_m\}\in \mathbb{R}^{m\times m}$ and $\boldsymbol{\Sigma}_c \in \mathbb{R}^{m\times m}$ represent the covariance matrices. Thus, the Kullback-Leibler divergence term can be simplified as

\begin{align} \label{eq:kl_simplify}
D_{KL}(p_{\varphi}(\cdot)||r(\cdot))=&\frac{1}{2}[\left(\boldsymbol{\mu}-\boldsymbol{\mu}_c\right)^T \boldsymbol{\Sigma}_c^{-1}\left(\boldsymbol{\mu}-\boldsymbol{\mu}_c\right)\\
    &-\log \operatorname{det}\left(\boldsymbol{\Sigma}_c^{-1} \boldsymbol{\Sigma}\right)+\operatorname{Tr}\left(\boldsymbol{\Sigma}_c^{-1} \boldsymbol{\Sigma}\right)-m].\notag
\end{align}

Instead of assigning predefined conditional priors for $r(\mathbf{\hat{z}}|y)$~\cite{guo2021conditional}, we choose to assign the conditional priors based on the received latent of the previous epoch. This means that if we denote $\boldsymbol{\mu}'_{c,t-1}=\mathbb{E}[\mathbf{\hat{Z}}|y=c]$, and $\boldsymbol{\Sigma}'_{c,t-1}=\mathbb{E}[\mathbf{\hat{Z}\hat{Z}^T}|y=c]-\boldsymbol{\mu}'_{c,t-1}\boldsymbol{\mu}'^T_{c,t-1}$ as the class conditional mean and covariance from the $(t-1)$-th epoch, then in the $t$-th epoch, the conditional latent prior for class $c$ is $r_t(\mathbf{\hat{z}}|y=c)=\mathcal{N}(\mathbf{\hat{z}}|\boldsymbol{\mu}'_{c,t-1},\boldsymbol{\Sigma}'_{c,t-1})$.

Furthermore, by taking into account the facts that the variational prior depends on the label information $Y$ and $H(Y)$ is a constant, we can derive an unbiased backpropagation-friendly upper bound estimation using the re-parameterization trick and Monte Carlo sampling~\cite{alemi2016deep},
\begin{equation}\label{eq:monte}
\begin{aligned}
    \mathcal{L}_{VCCIB}(\varphi,\psi)\simeq& \frac{1}{N}\sum_{n=1}^N(\beta D_{KL}(p_{\varphi}(\mathbf{\hat{z}}|\mathbf{x}_n)||r(\mathbf{\hat{z}}|y_n))\\
    &-\mathbb{E}_{\mathbf{\hat{Z}}}[\log q_{\psi}(y_n|\mathbf{\hat{z}})]).
\end{aligned}
\end{equation}
\begin{figure*}[ht]
    \centering
    \begin{subfigure}{0.40\linewidth}\label{fig:tsneJSCC}
        \centering
        \includegraphics[width=\linewidth]{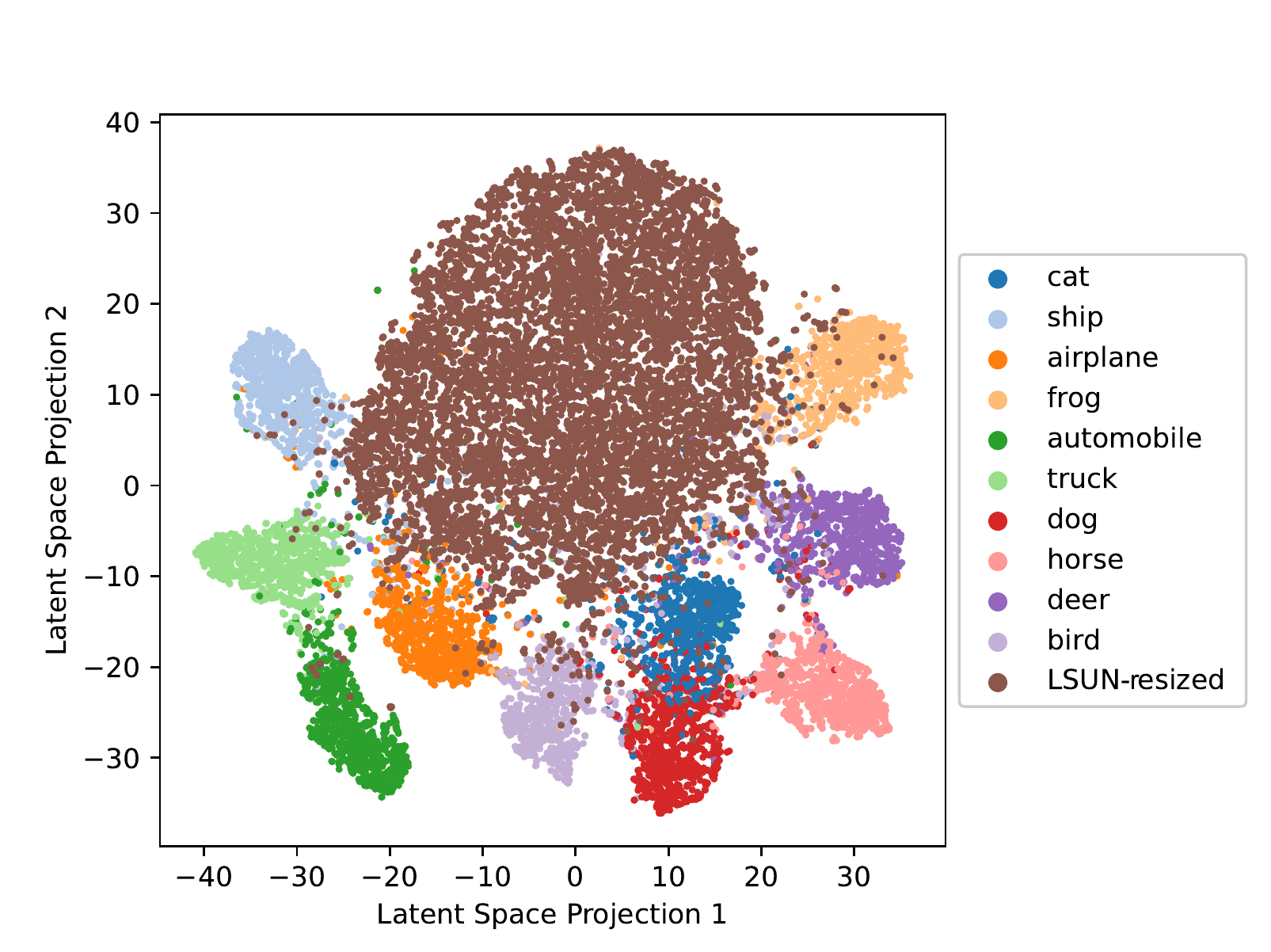}
        \caption{DeepJSCC: AUROC = 0.9314, PSNR = 20 dB}
    \end{subfigure}
    \begin{subfigure}{0.40\linewidth}\label{fig:tsneproposed}
        \centering
        \includegraphics[width=\linewidth]{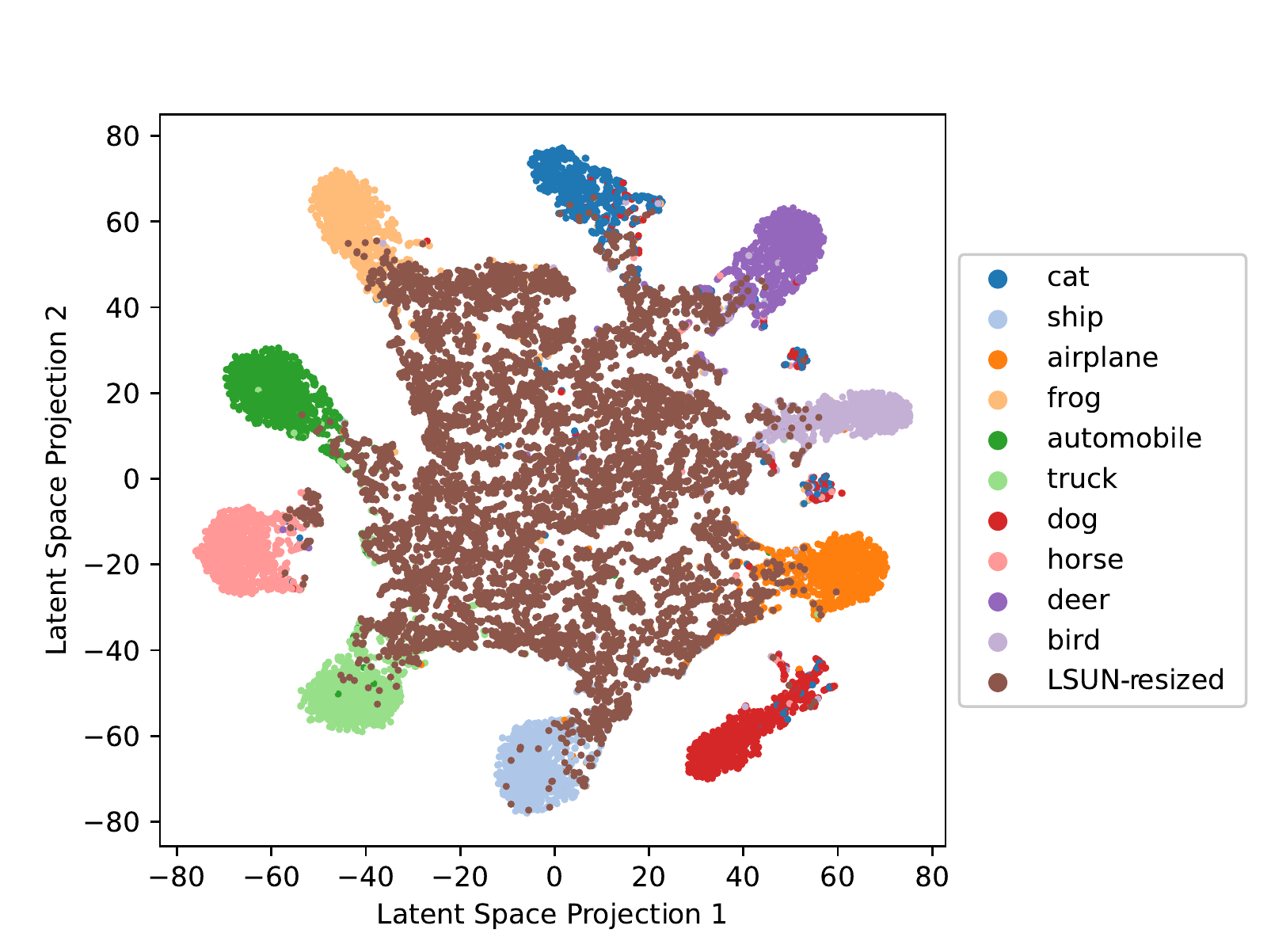}
        \caption{VCCIB: AUROC = 0.9657, PSNR = 20 dB}
    \end{subfigure}
    \caption{\textit{T}-SNE latent space projection for ID data CIFAR10 and OoD data LSUN-resized in the classification task with PSNR = 20 dB, classification accuracy $\geq 93\%$ and latency $\leq$ 6.5\text{ms.}}
    \label{fig:tsne}
\end{figure*}
\begin{figure*}[h]
    \centering
    \begin{subfigure}{0.40\linewidth}
        \centering
        \includegraphics[width=\linewidth]{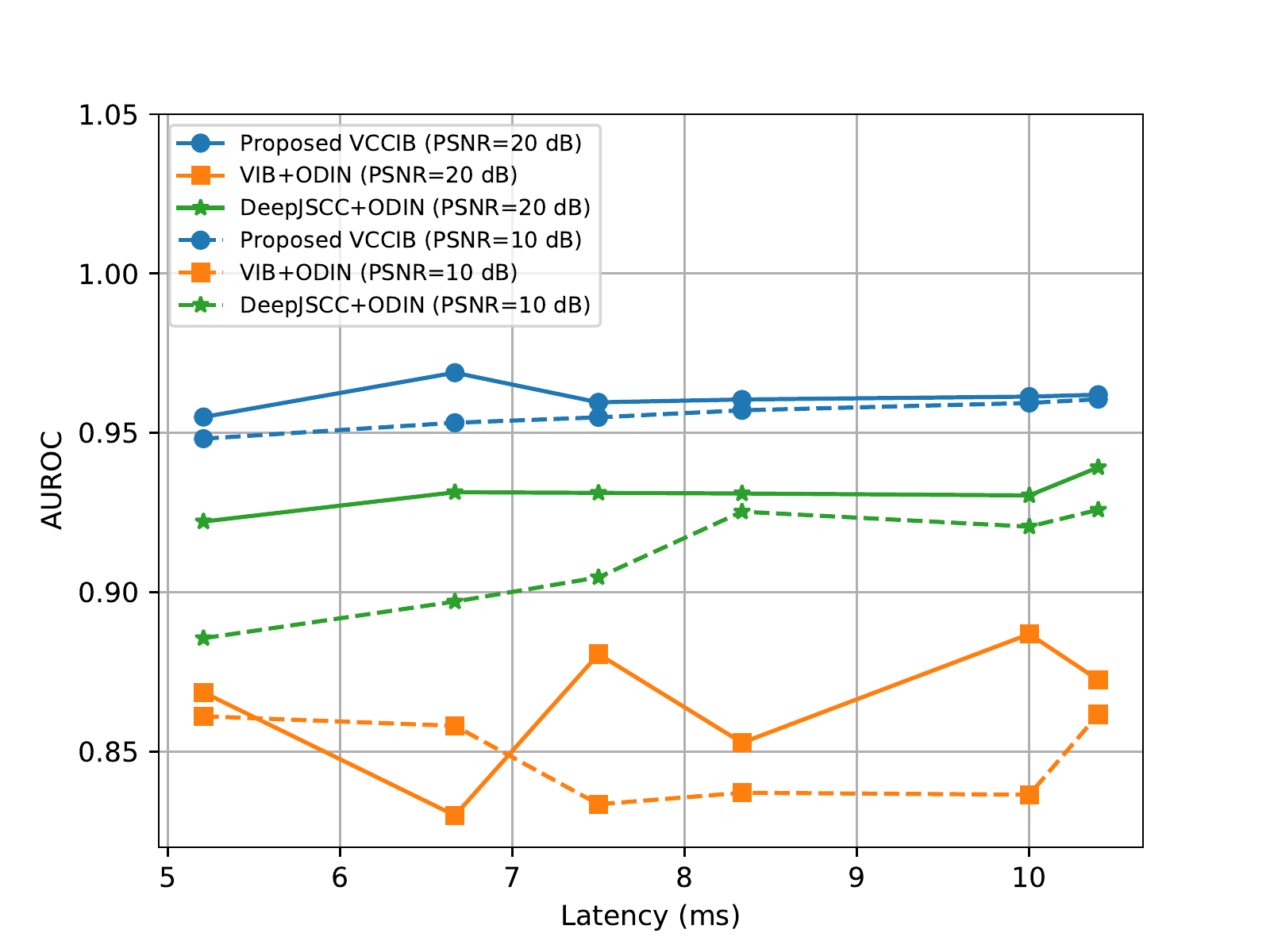}
        \caption{Unknown data: LSUN-resized}
        \label{fig:static_lsun}
    \end{subfigure}
    \begin{subfigure}{0.40\linewidth}
        \centering
        \includegraphics[width=\linewidth]{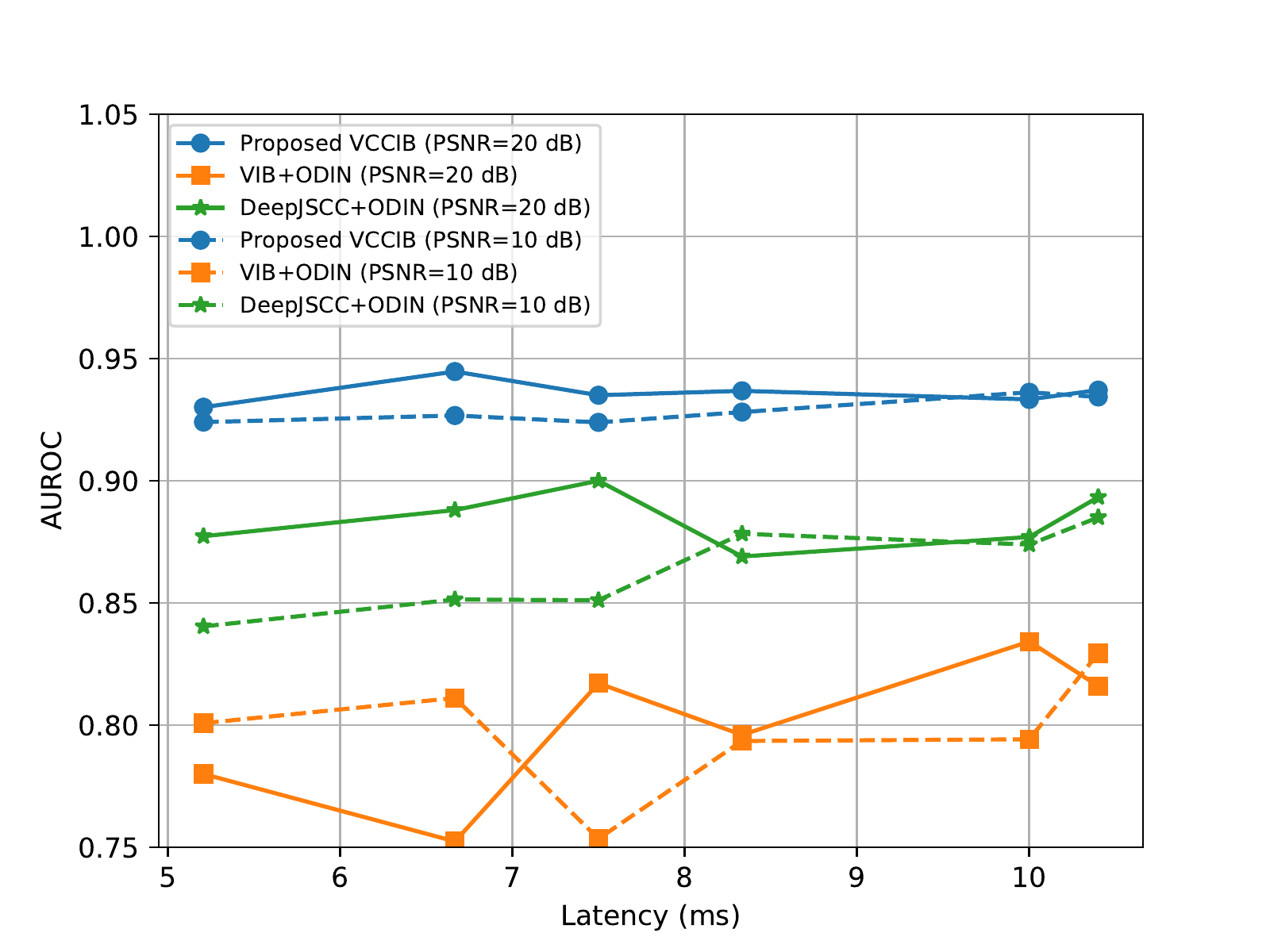}
        \caption{Unknown data: ImageNet-resized}
        \label{fig:static_imagenet}
    \end{subfigure}
    \caption{The AUROC-latency curves in static channel conditions with CIFAR10 as training data}
    \label{fig:static}
\end{figure*}

\begin{figure*}[h]
    \centering
    \begin{subfigure}{0.40\linewidth}
        \centering
        \includegraphics[width=\linewidth]{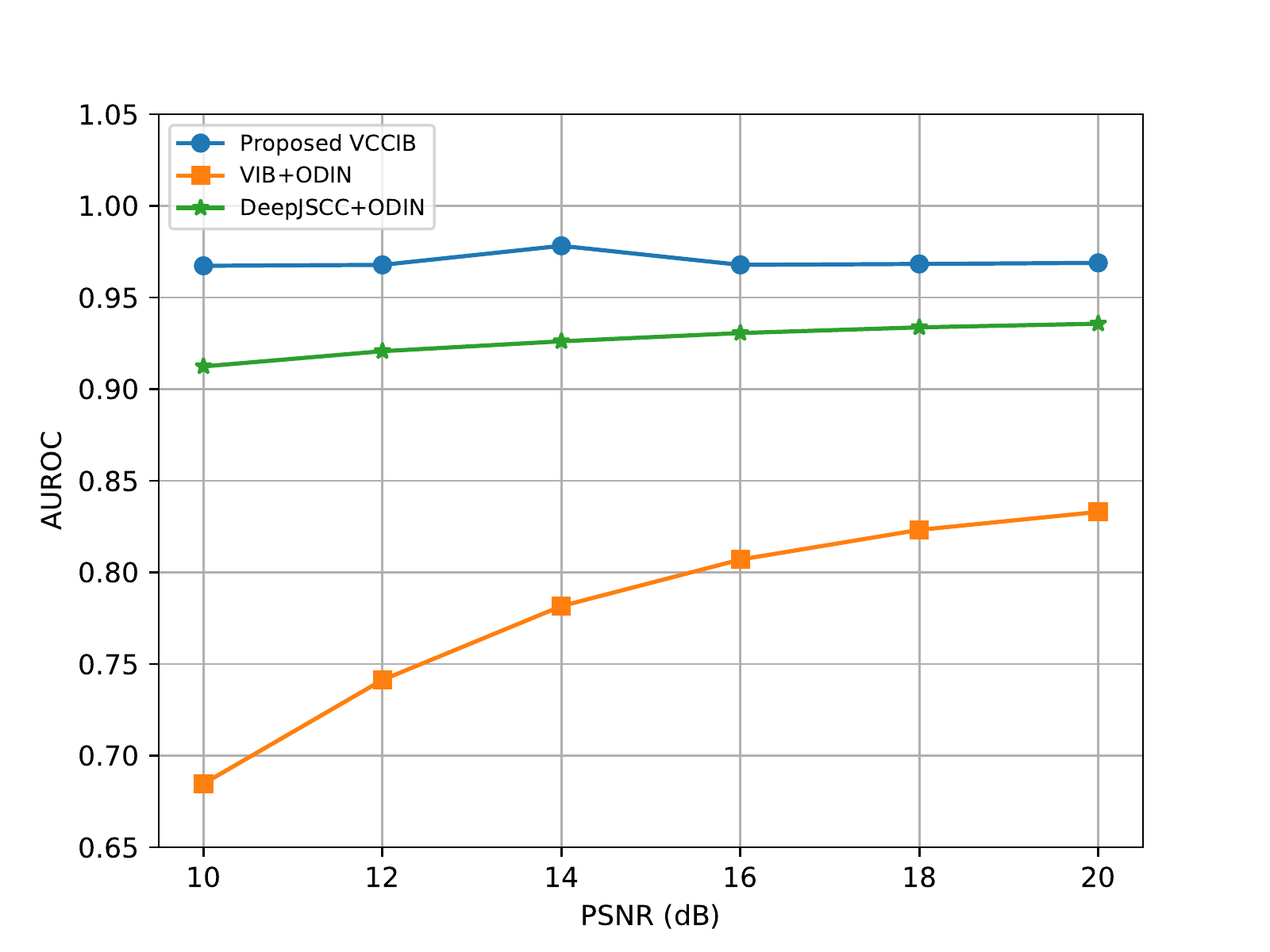}
        \caption{Unknown data: LSUN-resized}
        \label{fig:dynamic_lsun}
    \end{subfigure}
    \begin{subfigure}{0.4\linewidth}
        \centering
        \includegraphics[width=\linewidth]{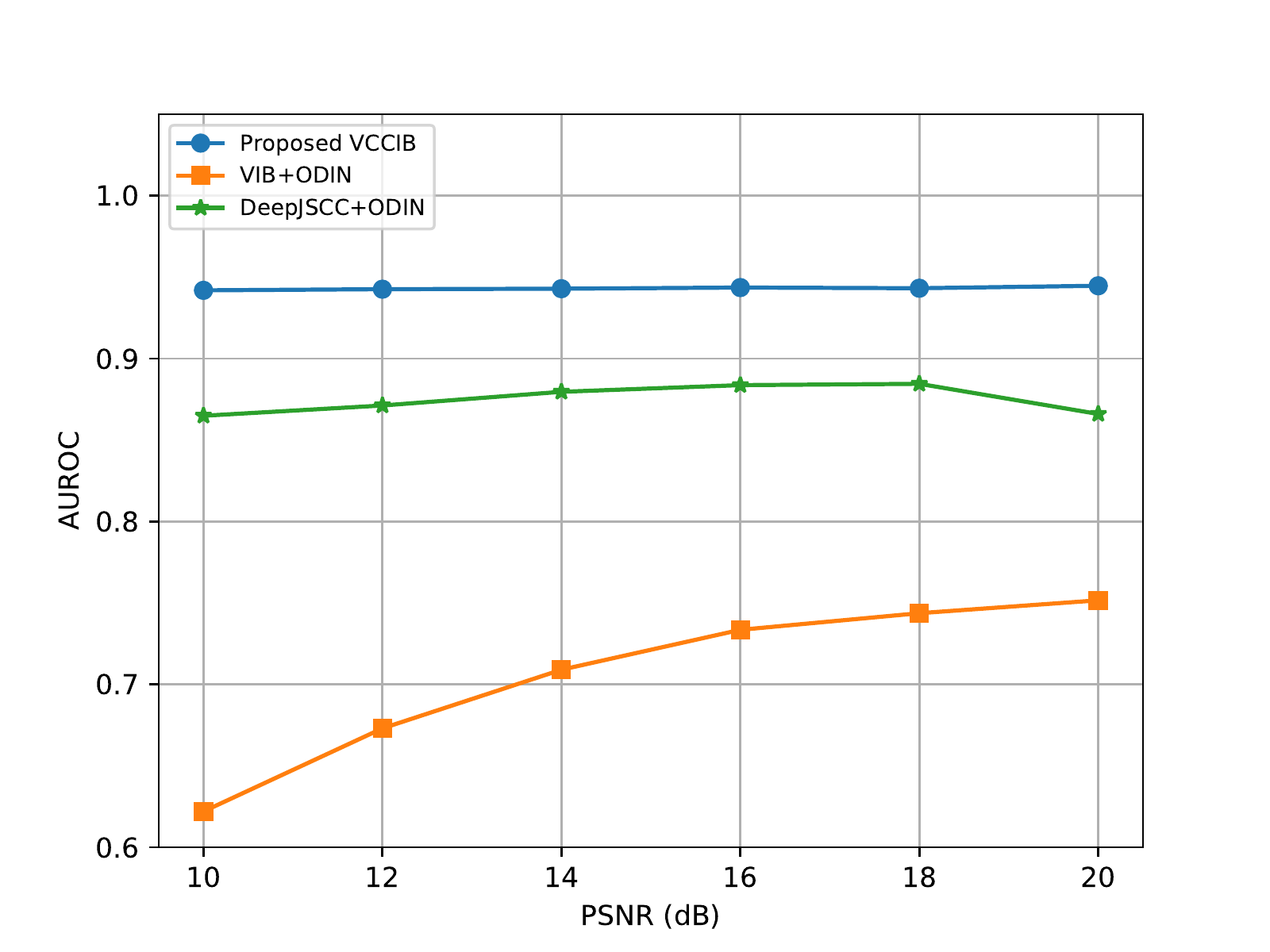}
        \caption{Unknown data: ImageNet-resized}
        \label{fig:dynamic_imagenet}
    \end{subfigure}
    \caption{The AUROC performance curves in dynamic channel conditions with CIFAR10 as training data}
    \label{fig:dynamic}
\end{figure*}

As shown by the first Kullback-Leibler divergence term in objective (\ref{eq:monte}), the CCIB approach aims to approximate the unique latent prior distribution corresponding to each class of data. To effectively detect the OoD data, these latent priors should be far away from each other, and the OoD data should have low and approximately equal likelihood with respect to each ID class~\cite{dinari2022variational}. Motivated by this idea, we use a triplet loss $\mathcal{L}_t$ given by
\begin{equation}\label{eq:triplet loss}
\begin{aligned}
    \mathcal{L}_t =& \frac{1}{T}\frac{1}{C}\sum_{c=1}^{C}\sum_{t=1}^{T}\max(||\mathbf{\hat{z}}_{r,c,t}
    -\mathbf{\hat{z}}_{m,c,t}||^2_2\\
    &-||\mathbf{\hat{z}}_{r,c,t}-\mathbf{\hat{z}}_{n,-c,t}||^2_2+\alpha,0),
\end{aligned}
\end{equation}
where $T$ is the total number of triads contained in $\{\mathbf{x},y\}_{i=1}^N$ and $C$ is the number of ID classes. This method is popular in contrastive learning~\cite{schroff2015facenet} to encourage separation among the latent representations of different labels. The triplet loss requires a reference latent representation $\mathbf{\hat{z}}_{r,c}$ and a matching latent representation $\mathbf{\hat{z}}_{m,c}$ from class $c$, as well as a non-matching latent representation $\mathbf{\hat{z}}_{n,-c}$ from another class and a predefined margin $\alpha$. The triplet loss in (\ref{eq:triplet loss}) aims to make the positive pair $(\mathbf{\hat{z}}_{m,c},\mathbf{\hat{z}}_{r,c})$ closer and the negative pair $(\mathbf{\hat{z}}_{m,c},\mathbf{\hat{z}}_{n,-c})$ further apart.

Finally, we have the contrastive learning-based VCCIB objective $\mathcal{L}'_{VCCIB}$, given by

\begin{align}\label{eq:VCCIB-c}
        \mathcal{L}'_{VCCIB} =& \frac{1}{N}\sum_{n=1}^N(\beta D_{KL}(p_{\varphi}(\mathbf{\hat{z}}|\mathbf{x}_n)||r(\mathbf{\hat{z}}|y_n))\\
    &-\mathbb{E}_{\mathbf{\hat{Z}}}[\log q_{\psi}(y_n|\mathbf{\hat{z}})])+\frac{1}{T}\frac{1}{C}\sum_{c=1}^{C}\sum_{t=1}^{T}\max(||\mathbf{\hat{z}}_{r,c,t}\notag \\
    &-\mathbf{\hat{z}}_{m,c,t}||^2_2-||\mathbf{\hat{z}}_{r,c,t}-\mathbf{\hat{z}}_{n,-c,t}||^2_2+\alpha,0).\notag
\end{align}
During the training phase, we train the proposed task-oriented communication system in an end-to-end paradigm with the objective (\ref{eq:VCCIB-c}) and the source domain dataset $\{\mathbf{{x}},y\}^N_{n=1}$. The edge server will store the parameters of conditional Gaussian prior $\{\boldsymbol{\mu_{T,c},\boldsymbol{\Sigma_{T,c}}}\}^C_{c=1}$ as well as the mean and covariance matrix of the intermediate features for all $C$ classes data after the $T$-th training epoch. In the testing phase, given a data point $\mathbf{x}$, the proposed task-oriented communication system will calculate the log-likelihood score $\{S_c(\mathbf{x};\varphi,\psi)\}^C_{c=1}$ of each ID class based on the stored Gaussian distributions. To enhance the performance of the OoD detection, we use test-time reconstruction-based weighting for the log-likelihood score. We assume that the edge device is equipped with an auto-encoder trained on the source domain data. Given a test data point $\mathbf{x}$, the on-device auto-encoder outputs a reconstructed data $\mathbf{\hat{x}}$ in parallel and calculates the MSE loss $l_{\mathbf{x}}=(\mathbf{x}-\mathbf{\hat{x}})^2$. Then the weighted log-likelihood score can be expressed as  $\{S'_c(\mathbf{x};\varphi,\psi)|S'_c(\mathbf{x};\varphi,\psi)=l_{\mathbf{x}}S'_c(\mathbf{x};\varphi,\psi),\forall c\in C\}$. The log-likelihood based OoD detection function $J(\mathbf{x};\varphi,\psi,\lambda)$ at the decoder is given by
\begin{equation}
\small
    J(\mathbf{x};\varphi,\psi,\lambda)= 
    \begin{cases}\text {ID data, class $k$,}\text { if }k=\operatorname*{argmax}_c S'_c, \\ 
    \quad\quad\quad\quad\quad\quad\quad\text{and } S'_k\geq\lambda, \\ 
    
    \text {OoD data,}\text { if }k=\operatorname*{argmax}_c\ S'_c,\\
    \quad\quad\quad\quad\quad\quad\quad\text{and } S'_k<\lambda. \\
    \end{cases}
\end{equation}



\section{Performance Evaluation}
\subsection{Experiments Setup}
\textit{1) Basic Setup and Metrics:} We choose the CIFAR-10 dataset as the ID dataset, and the LSUN-resized dataset and Tiny ImageNet-resized dataset as the OoD dataset. For fairness, we use ResNet-18 as the backbone network in all experiments. Considering the transmit power limitation, we use the $sigmoid(\cdot)$ function as the activation function in the last layer of the JSCC encoder. Furthermore, we use the same communication setting as~\cite{shao2021learning} where the symbol rate is 9600 Baud\footnote{The code is available at {github.com/hlidmhkust/VCCIB}.}.

For the OoD detection in the classification task, we use the area under the receiver operating characteristic curve (AUROC) to indicate the detection performance. AUROC measures the ability of a model to distinguish between positive and negative classes by calculating the area under the receiver operating characteristic (ROC) curve, where ROC is a plot of the true positive rate against the false positive rate at various classification thresholds.

\textit{2) Baselines}: We compare the proposed method with DeepJSCC and VIB schemes:
\begin{itemize}
    \item \textbf{DeepJSCC}: An end-to-end DL-based JSCC scheme with the cross-entropy loss function~\cite{bourtsoulatze2019deep}.
    \item \textbf{VIB}: A variational information bottleneck framework with the cross-entropy and Kullback-Leibler divergence loss function~\cite{shao2021learning}. Here, we assume the latent prior is a standard multivariate Gaussian distribution, i.e., $\mathbf{z}\sim\mathcal{N}(0,\mathbf{I})$.
\end{itemize}
We combine these two methods with the popular OoD detection tool ODIN~\cite{liang2017enhancing}. The main features of ODIN include the $\gamma$ hyperparameter scaling and input pre-processing. Following~\cite{hsu2020generalized}, we set $\gamma=1000$ for a fair comparison.

\subsection{Experiments Results}
In Fig. \ref{fig:tsne}, we present the \textit{t}-distribution stochastic neighbor embedding (\textit{t}-SNE) of the noise-corrupted features combined with intermediate features in the receiver of both CCIB and DeepJSCC. As shown in the figure, it is difficult to distinguish the intermediate features of the ID data and the OoD data by the DeepJSCC method. On the contrary, we can observe that, with the CCIB-based model, the ID data is equally away from the OoD data, which indicates the CCIB-based model has a higher entropy on the OoD data. This is attributed to the use of a conditional prior and a contrastive learning-based separation loss, which enforces the separation of the latent representations from different labels. By doing so, the CCIB-based approach can effectively differentiate ID and OoD data, while maintaining the compactness and informativeness of the extracted features.
\begin{figure*}[h]

\begin{equation}\label{eq:proof}
\scriptstyle
    \begin{aligned}
        \mathcal{L}_{CCIB}=& \beta\int_{\mathcal{X,Y,\hat{Z}}}p(\mathbf{x},y,\mathbf{\hat{z}})\log \frac{p(\mathbf{x},\mathbf{\hat{z}}|y)}{p(\mathbf{x}|y)p(\mathbf{\hat{z}}|y)}d\mathbf{x}dyd\mathbf{\hat{z}}-\int_{\mathcal{X,Y,\hat{Z}}}p(\mathbf{x},y,\mathbf{\hat{z}})\log\frac{p(y,\mathbf{\hat{z}})}{p(y)p(\mathbf{\hat{z}})}d\mathbf{x}dyd\mathbf{\hat{z}}\\
        \leq& \beta\int_{\mathcal{X,Y,\hat{Z}}}p(\mathbf{x},y)p_{\varphi}(\mathbf{\hat{z}}|\mathbf{x})\log\frac{p_{\varphi}(\mathbf{\hat{z}}|\mathbf{x})}{r(\mathbf{\hat{z}}|y)}     d\mathbf{x}dyd\mathbf{\hat{z}}-\int_{\mathcal{X,Y,\hat{Z}}}  p(\mathbf{x},y)p_{\varphi}(\mathbf{\hat{z}}|\mathbf{x})\log q_{\psi}(y|\mathbf{\hat{z}})  d\mathbf{x}dyd\mathbf{\hat{z}}-\underbrace{ H(Y)}_{constant}\\
        =&\beta\int_{\mathcal{X,Y,\hat{Z}}}p(\mathbf{x},y)D_{KL}(p_{\varphi}(\mathbf{\hat{z}}|\mathbf{x})||r(\mathbf{\hat{z}}|y))d\mathbf{x}dyd\mathbf{\hat{z}}-\int_{\mathcal{X,Y,\hat{Z}}}  p(\mathbf{x},y)p_{\varphi}(\mathbf{\hat{z}}|\mathbf{x})\log q_{\psi}(y|\mathbf{\hat{z}})  d\mathbf{x}dyd\mathbf{\hat{z}}+\text{constant}.
    \end{aligned}
\end{equation}
\hrule
\end{figure*}
In Fig. \ref{fig:static}, we present the AUROC metric evaluated at different latencies with PSNR $=10$ dB and PSNR $=20$ dB on the CIFAR10 image classification tasks. Here, the latency is proportional to the transmitted feature dimensions, and the PSNR is kept the same for training and testing. As shown in Fig. \ref{fig:static}, the proposed CCIB-based method significantly outperforms the two baselines when evaluated on the LSUN-resized and Tiny ImageNet-resized OoD datasets. The VIB method achieves the worst AUROC on the OoD detection performance because it assumes a fixed Gaussian prior distribution for all input data, which makes VIB insensitive to the distribution of the input data during testing. On the other hand, the CCIB-based method achieves the best AUROC on the OoD detection performance by assuming different conditional priors, which leads to the separation of latent representations and makes the system more sensitive to the OoD data.

To evaluate the OoD detection performance in dynamic channel conditions, we set the transmitted feature dimension to $64$, and the training PSNR to $20$ dB for all methods. In Fig. \ref{fig:dynamic}, we present the OoD detection performance of these three methods with different testing PSNR. As shown in the figure, the proposed CCIB method outperforms the two baselines. Furthermore, the proposed CCIB-based scheme and DeepJSCC scheme are almost unchanged when evaluated on both the LSUN-resized and Tiny ImageNet-resized OoD datasets. On the contrary, VIB has better AUROC performance in the high PSNR regime because it can alleviate the degree of confusion due to the fixed latent prior in high PSNR regime.
\section{Conclusions}

In this paper, we proposed a task-oriented communication system based on CCIB for reliable edge-assisted inference with OoD detection. The proposed CCIB approach builds upon the IB framework by incorporating the conditional prior and contrastive learning methods. It can enhance the distinguishability of ID latent representations compared to unknown OoD representations, while maintaining their informativeness and compactness. We demonstrated the effectiveness of the proposed approach for OoD detection on an image-based task-oriented communication system. In the future, it is important to further reduce the latency while maintaining the robustness for OoD detection.

\appendices
\section{Proof of Theorem 1}
\label{FirstAppendix}

According to the non-negativity of Kullback-Leibler divergence, if we use the variational distribution $q(y|\mathbf{\hat{z}})$, $r(\mathbf{\hat{z}}|y)$ to approximate $p(y|\mathbf{\hat{z}})$ and the true conditional latent prior $p(\mathbf{\hat{z}}|y)$, we can derive the variational upper bound of $\mathcal{L}_{CCIB}$ in (\ref{eq:proof}).

\linespread{0.96}{
\bibliographystyle{IEEEtran}
\bibliography{IEEEabrv,ref}
}



\end{document}